\documentclass[twocolumn]{book}
\usepackage{graphicx,color}
\usepackage{makeidx,universe}

\makeauthorindex
\BookTitle{Proceedings of the XXIX PHYSICS IN COLLISION}

\CopyRight{\copyright 2009 by Universal Academy Press, Inc.}

\begin{document} %******************************************

%\tableofcontents
\pagenumbering{arabic}

\chapter{%
{\LARGE \sf
Heavy-Ion Collisions - Examining the Quark Gluon Plasma at RHIC} \\
{\normalsize \bf %%%%%%%%%%%%%%******* Authors **************
Helen Caines,$^{1}$ } \\
{\small \it \vspace{-.5\baselineskip}%***** Affiliations ***********
(1) Physics Department, Yale University, New Haven, CT 06520, U.S.A
}
}

%**************************
% Please note:
% One \AuthorContents{} is necessary
% for EACH CONTRIBUTION, for the contents page and
% One \AuthorIndex{} is necessary
% for EACH AUTHOR, for the index.
%**************************

%***** Item below is the data for CONTENTS. 
%***** Please enter all author's name that should be initialized.
\AuthorContents{H.\ Caines} 

%***** items below are the data for AUTHOR INDEX. 
%***** Please enter a author's name that should be initialized.
\AuthorIndex{Caines}{H.} 

  \baselineskip=10pt %*******
  \parindent=10pt    %*******

\section*{Abstract} %******** Body of document starts.****************

The main goals of relativistic heavy-ion experiments is to study the properties of QCD matter  under extreme temperatures and densities.   The focus of this talk is the studies that are underway at the Relativistic Heavy Ion Collider (RHIC), located at the Brookhaven National Laboratory (BNL) on Long Island, New York, U.S.A. I discuss selected highlights from the past couple of years that are key to elucidating the characteristics  of the new state of matter created in these heavy-ion collisions, called a Quark Gluon Plasma.
\section{Introduction}

The goals of the experiments at the Relativistic Heavy Ion Collider (RHIC) are to study hot QCD matter, and investigate the  spin structure of the proton.  RHIC has now been running for nearly a decade, and has collided a diverse set of beams from p to Au ions at $\sqrt{s_{_{NN}}}$ = 20-200 GeV (p beams have just been collided at  $\sqrt{s}$ = 500 GeV). Initially there were four RHIC experiments BRAHMS~\cite{Brahms}, PHENIX~\cite{Phenix}, PHOBOS~\cite{Phobos},  and STAR~\cite{Star}. Only the  two large multi-purpose experiments, PHENIX and STAR, remain in operation, BRAHMS and PHOBOS completed their programs and have been de-commisioned. This talk focuses on recent results from these two collaborations, which  highlight  our  progress in probing the properties of  the new state of matter, called a Quark Gluon Plasma (QGP),  created  in relativistic heavy-ion collisions. 

At sufficiently high temperatures it has long been predicted that strongly interacting matter, described by QCD, must ``melt" into a deconfined phase of quarks and gluons~\cite{QGPPredict}.  Lattice calculations indicate that at non-zero baryo-chemical potential the critical temperature for such a transition is  T$_{C}$150-180 MeV~\cite{Lattice}. There is clear evidence for the RHIC experiments that such a deconfined phase, the QGP, is created in Au-Au collisions at $\sqrt{s_{_{NN}}}$ = 200 GeV. The initial energy density in the collision region is greater than 4.6 GeV/fm${4}$~\cite{EnergyDensity}  and the initial temperature is T=300-600 MeV~\cite{DirectPhoton}, which are much higher than that where hadrons can exist. The  degrees of freedom in the initial stages are those of quarks and gluons. Together these results show that a new partonic state of matter  exists in the early stages of these heavy-ion collisions and that this matter flows like an almost ``perfect" liquid ~\cite{WPBrahms, WPPhenix, WPPhobos, WPStar}.  An initially surprising result was that the medium is almost opaque to color charges passing through it. In deference to this discovery the plasma is now frequently referred to as the ``sQGP", or strongly coupled QGP.  The use of  high transverse momentum probes (jets) to elucidate the properties of the sQGP are the topic of the first section of my talk. The second section discusses the tantalizing suggestion of observation of local strong parity violation in heavy-ion collisions where partonic degrees of freedom exist. Finally, I briefly discuss two  exciting new avenues in relativistic heavy-ion studies that will  become available in the next couple of years.

\section{Jets as Probes of the sQGP}

At RHIC energies  high transverse momentum ($p_{T}$),  or Q$^{2}$, scatterings can only occur in the initial parton-parton interactions. This means that a) the integrated yield of hard processes must scale with the number of binary nucleon-nucleon interactions in the  collisions  (binary scaling) and b) in heavy-ion events the scattered partons must pass through the matter produced before they escape the medium and  fragment. Hence, we have a direct, and calibrated, probe of the partonic phase of the collision. These interactions with the medium will result in the attenuation and/or complete absorption of the energy of the jet into the bulk. By comparing the measured jet rates and properties, as a function of p$_{T}$, with those from p-p events at the same collision energy we can extract details about the sQGPs properties.

\subsection{Di-hadron Correlations}
\begin{figure}[h]
  \begin{center}
    \includegraphics[width=\linewidth]{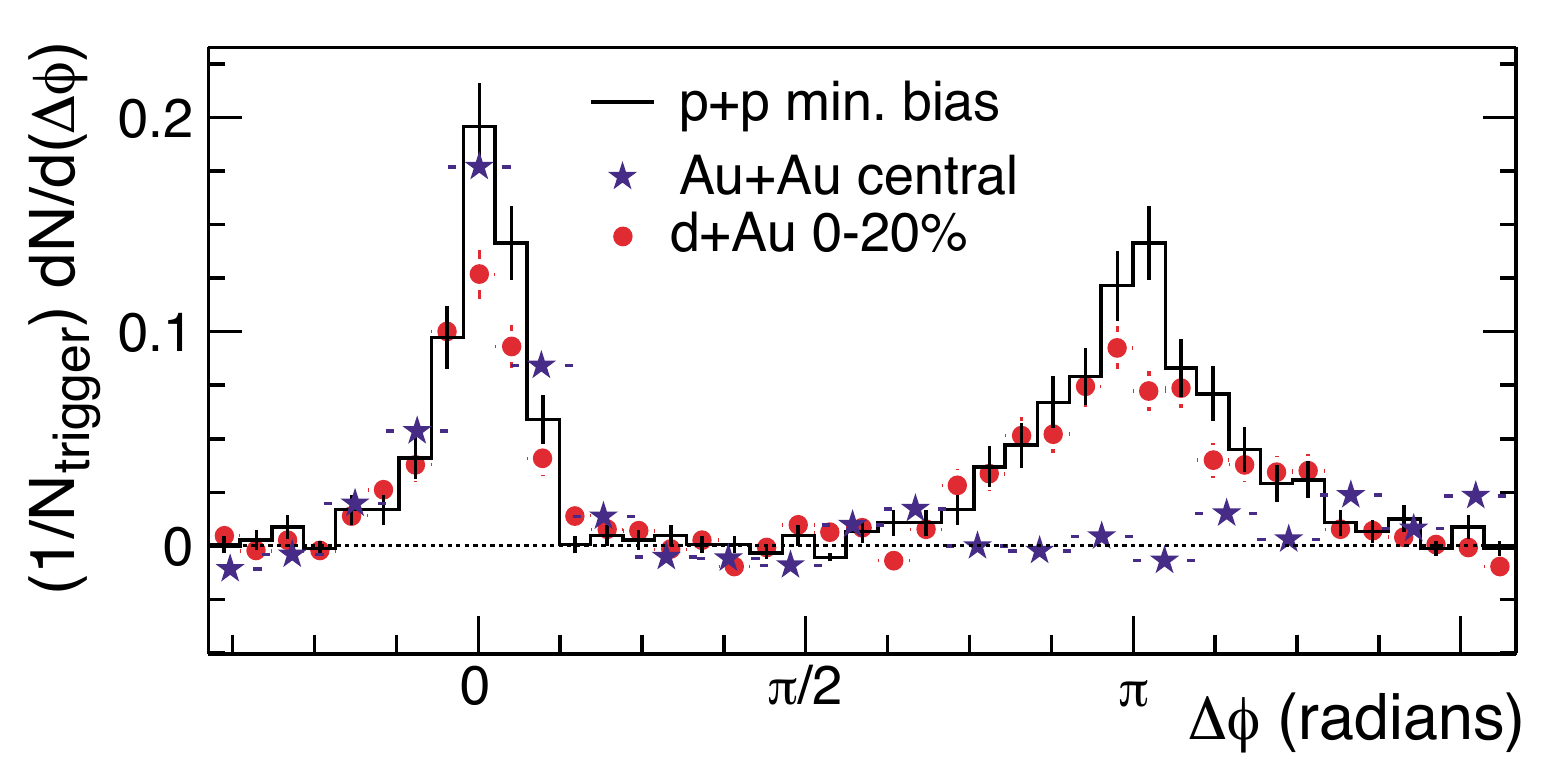}
  \end{center}
  \vspace{-1pc}
  \caption{Color Online:  Di-hadron $\Delta\phi$ correlations for p-p (solid histogram), 0-20$\%$ d-Au (red circles ) and central Au-Au (blue stars) collisions at $\sqrt{s_{_{NN}}}$ = 200 GeV. Figure from~\cite{dihadron}.} \label{Fig:Dihadron}
\end{figure}

Measuring jets in the heavy-ion environment is a highly non-trivial task. First studies  utilized high p$_{T}$ leading particle as proxies for full jet reconstruction and di-hadron correlations. A high p$_{T}$ ``trigger" particle is selected and the $\Delta\phi$  distribution of all other associated particles in the events constructed. In p-p collisions clear jet peaks are observed at  $\Delta\phi$=0 and $\pi$, (solid histogram in Fig.~\ref{Fig:Dihadron} ). However, while a clear peak is seen at $\Delta\phi$=0 in the central Au-Au data (blue stars), it has disappeared on the away side ($\Delta\phi$= $\pi$).  The presence of the away side peak in the d-Au data (red circles), where no QGP is created, shows that this jet quenching phenomena is not caused by initial of final state  cold matter effects  \cite{dihadron}. The quenching of the away-side jet in Au-Au collisions must therefore be due to  interactions of the partons with the sQGP. In addition, the  high p$_{T}$ particle yields in central Au-Au collisions have  been shown to be suppressed by a factor of five compared to binary scaled p-p yields \cite{HighPtSup}.

The suppression of single high p$_{T}$ particles and the quenching of the away-side jet  are both interpreted as dominantly due to multiple final-state gluon radiation off  the hard scattered partons, that is induced by the presence of the dense colored medium (the sQGP).  The magnitude of the  partonic energy loss depends on the length of the medium traversed and  the transport coefficient $\hat{q}$, which is the mean squared p$_{T}$ transferred  per unit length~\cite{qhat}.  Current data constrain  model-dependent calculations of $\langle \hat{q}  \rangle $  to within $\pm$20-30$\%$~\cite{PhenixQhat}. However the theoretical uncertainties are still  large and  estimates of $\langle \hat{q}  \rangle $ range from 5-15 GeV$^{2}$/fm.

Although much has been learned by studying single particle high p$_{T}$  yields and di-hadron correlations,  exact connections to pQCD calculations were limited  since they are only indirect measurements of jets. In principle, only by direct jet reconstruction can the initial parton energy  be fully recovered.

\subsection{Jet reconstruction}

\begin{figure}[h]
  \begin{center}
    \includegraphics[width=\linewidth]{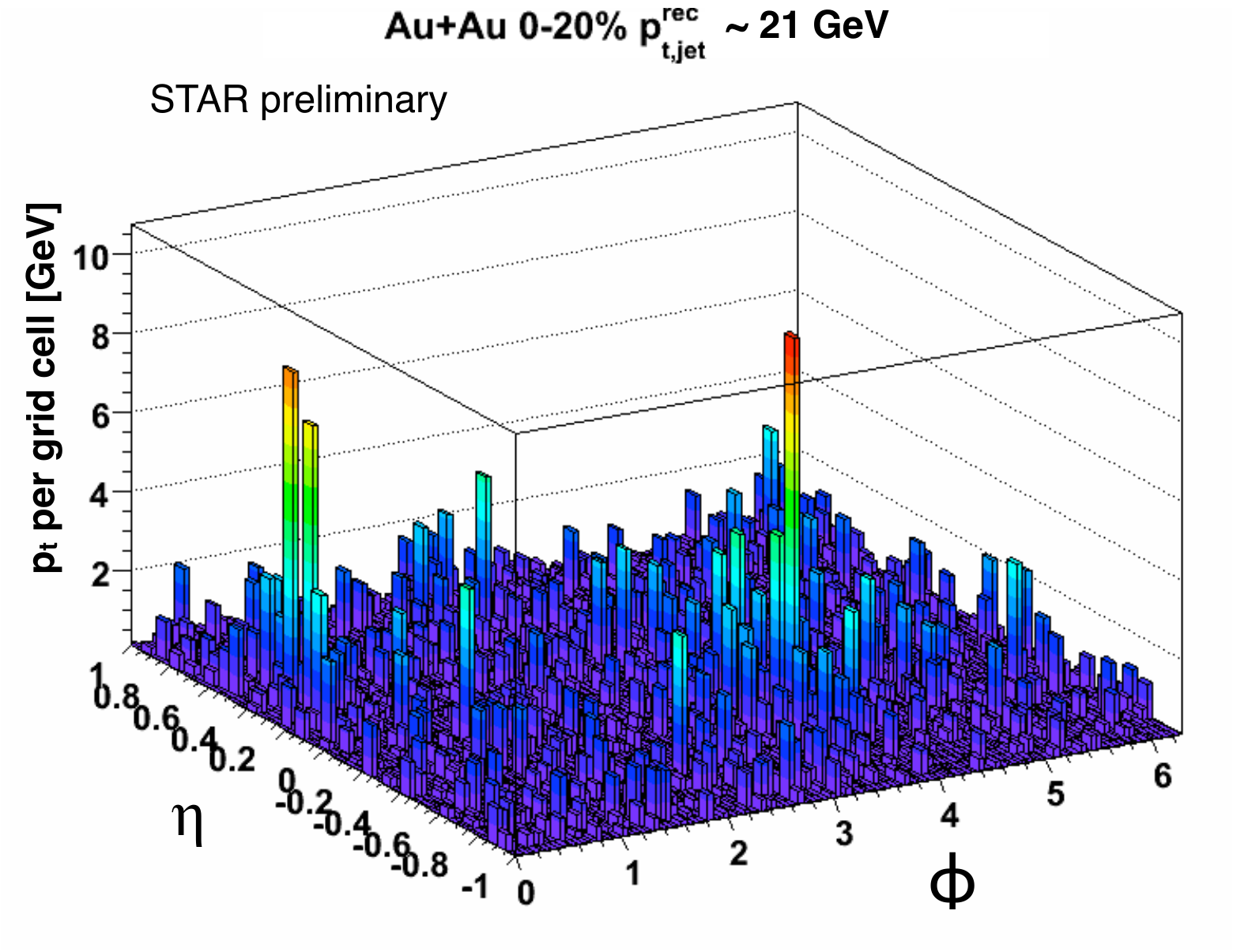}
  \end{center}
  \vspace{-1pc}
  \caption{Color Online:  The summed neutral and charged p$_{T}$ in a  0-20$\%$ central Au-Au collisions at  $\sqrt{s_{_{NN}}}$ = 200 GeV in $\eta-\phi$ space for an event with a reconstructed di-jet of p$_{T}^{rec} \approx$ 20 GeV/c. Figure from \cite{JoernHP}.} \label{Fig:STAREvent}
\end{figure}

The high multiplicity created in heavy-ion events means that all  jets are sat on a  large, and fluctuating background. However, as shown in Fig.~\ref{Fig:STAREvent} a jet of $\sim$ 20 GeV is still clearly visible above the background when the charged and neutral energy is taken into account, meaning that while the background cannot  be ignored, jet reconstruction in Au-Au events is possible. 

The two RHIC experiments currently utilize different jet finders. STAR uses a set of three different algorithms from the FastJet package~\cite{FastJet}, the  seedless cone algorithm (SISCone), and the k$_{T}$ and Anti-k$_{T}$  recombination codes. PHENIX meanwhile use a seedless Gaussian filter~\cite{Albrow:1979yc, GausFilt}. All of these jet finders are infrared and collinear safe. Both experiments use charged tracking detectors and electro-magnetic calorimetry at mid-rapidity to perform their studies

\begin{figure}[h] 
	\begin{center}
		\includegraphics[width=\linewidth]{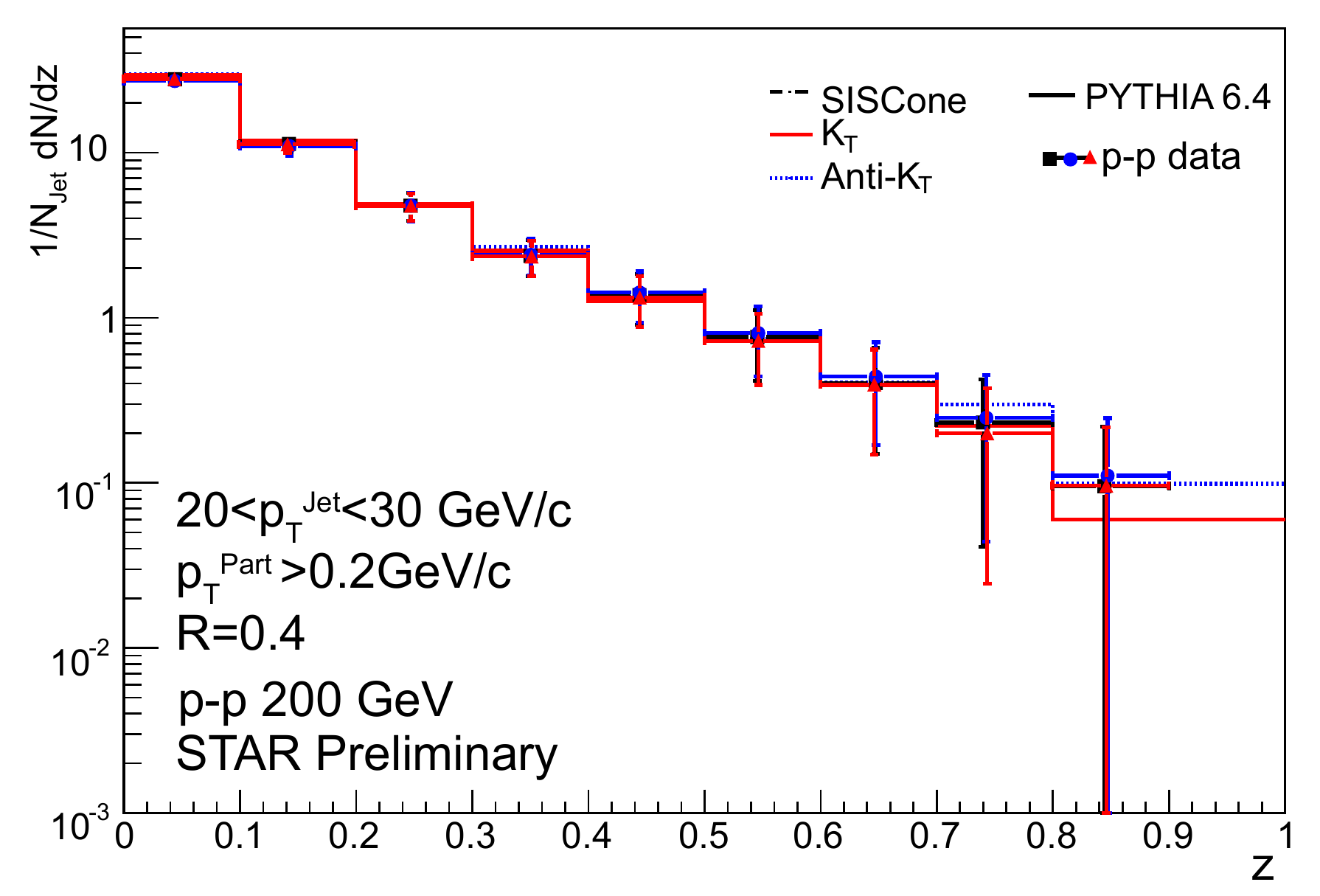}
		\includegraphics[width=\linewidth]{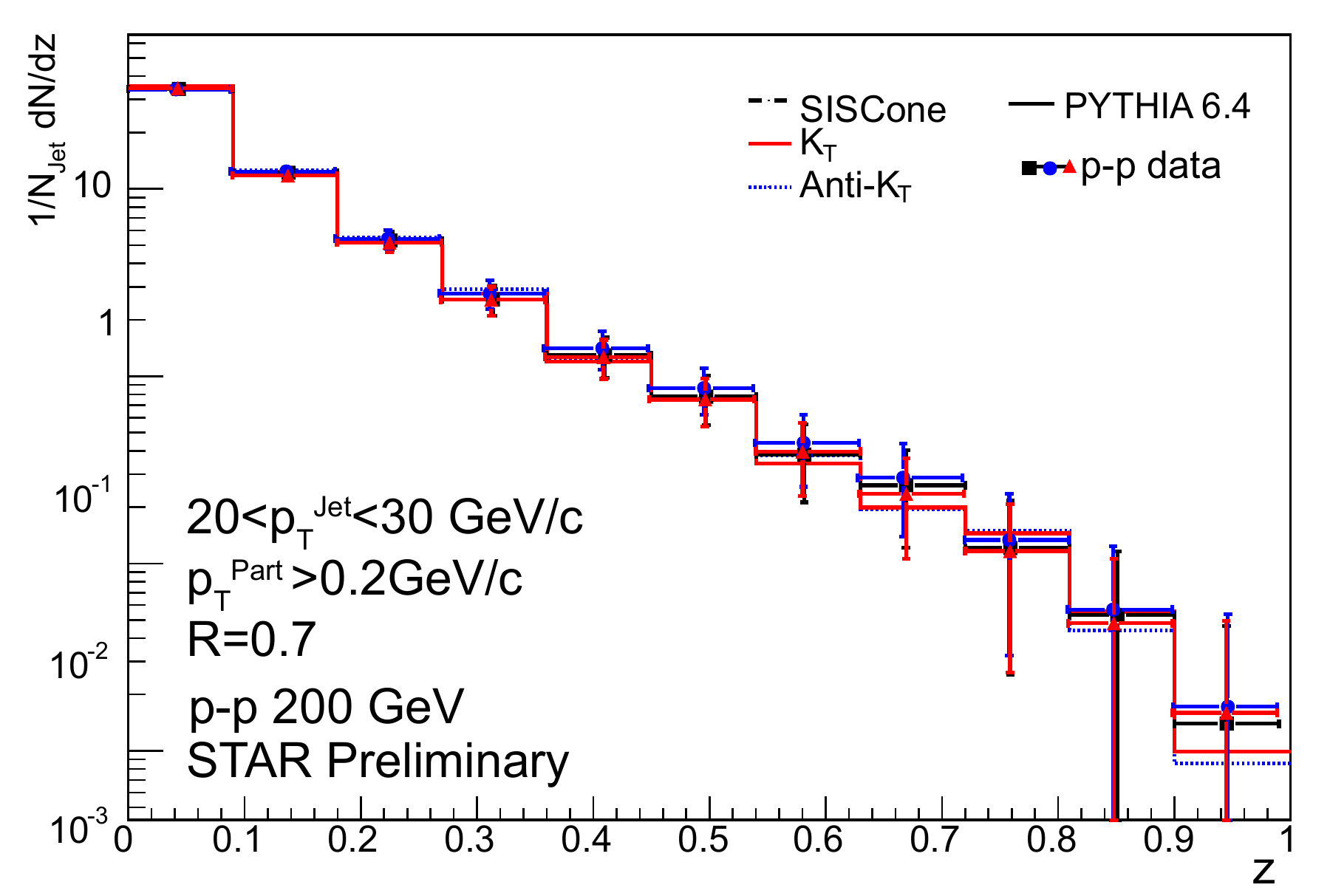}
		\caption{ Color online: Charged particle, detector level, $z$ fragmentation functions for jets reconstructed with 20$< p_T<$ 30 GeV/c  in $\sqrt{s}$ = 200 GeV p-p collisions, compared to PYTHIA for 3 different jet algorithms. $|\eta|<$1-R, for R=0.4 (top) and R=0.7 (bottom). Red triangles - k$_{T}$,  blue circles - Anti-k$_{T}$ and SISCone - black squares.  The errors  are statistical only. Figure from \cite{FFQM}.} \label{Fig:FFZ}
	\end{center}
\end{figure}

Before the  Au-Au analyses could proceed it was necessary to ensure that jet production in p-p collisions at  RHIC is a  calibrated probe that is well described by pQCD calculations. We find that the inclusive jet cross-section agrees with NLO predictions over seven orders of magnitude~\cite{JetCross}. The charged particle fragmentation functions have been extracted and are shown in  Fig.~\ref{Fig:FFZ}  for reconstructed jet p$_{T}$ of 20-30 GeV/c and resolution parameters of R=0.4 and 0.7. Charged particles with p$_{T}> $ 0.2 GeV/c and calorimeter towers with E$_{T} >$ 0.2 GeV are included in the study. These data have not yet been corrected  to the particle level, therefore the results are  compared to PYTHIA 6.410, Tune A~\cite{Pythia,PythiaTuneA}. These calculations passed through STAR's detector simulation and reconstruction algorithms. The single charged  particle reconstruction efficiency in the TPC is $>$ 80$\%$ for $p_{T}>$1 GeV/c, but drops steeply for lower $p_{T}$. Such detector inefficiencies  and  the presence of  particles which are undetected by the STAR detector, such as the neutron and K$^{0}_{L}$, cause the  reconstructed jet $p_{T}$  to be lower on average than the  true value. Good agreement is seen between PYTHIA and the data for jet algorithm resolution parameters R=0.4 and 0.7  suggesting that there are no significant large angle initial/final state radiation effects beyond those accounted for by this LO calculation. Data from studies of the underlying event in p-p events by STAR also support this conclusion~\cite{FFDPF}.

\begin{figure}[htb] 
	\begin{center}
		\includegraphics[width=\linewidth]{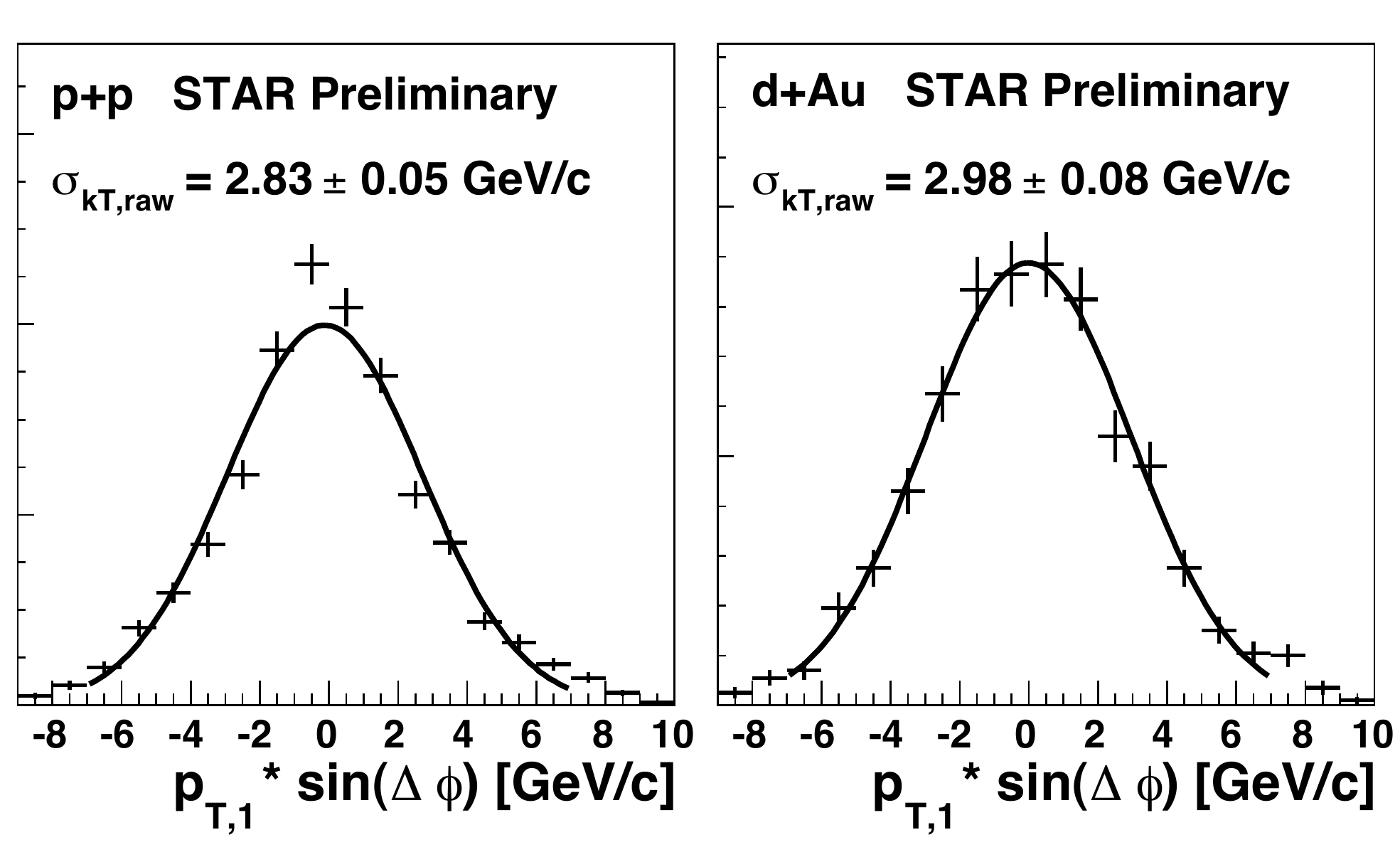}
		\caption{ Color online: Measured k$_{T,raw}$ distributions from p-p collisions (left) and  d-Au (right) $\sqrt{s_{_{NN}}}$ = 200 GeV collisions for 10 GeV/c$<$p$_{T,2}$$<$ 20 GeV/c. The solid curves are Gaussian fits to the data. Figure from \cite{JetKt}.} \label{Fig:JetKt}
	\end{center}
\end{figure}

Once again  initial state effects on jet properties need to be disentangled from  those of the medium. Di-hadron correlations  showed that the away-side signal was not suppressed in d-Au collisions but the affect of nuclear k$_{T}$  remained to be determined. The intrinsic k$_{T} $ of the nucleons in the nuclei and the  additional scattering of the parton in matter before fragmentation results in the scattered parton pairs no longer being exactly back-to-back in $\Delta\phi$. This effect can be measured as a broadening of  the $\Delta\phi$ di-jet correlation  in d-Au compared to p-p collisions. We therefore measure k$_{T,raw}$ = p$_{T,1}$ sin $(\Delta\phi)$, where  p$_{T,1} $ is the  highest jet p$_{T}$ of the di-jet pair and $\Delta\phi$ is the angle between the jets.  Figure~\ref{Fig:JetKt} shows the measured k$_{T,raw}$  distributions from p-p collisions in the left plot and  d-Au collisions in the right plot for 10 GeV/c$<$p$_{T,2}$$<$ 20 GeV/c. The extracted $\sigma_{k_{T,raw}}^{p-p}$ = 2.8 $\pm$ 0.1 (stat) GeV/c and  $\sigma_{k_{T,raw}}^{d-Au}$ = 3.0 $\pm$ 0.1 (stat) GeV/c from Gaussian fits to the data indicate that any additional broadening due to nuclear effects is small.

Much work has been invested in developing codes and techniques that will allow us to extract jets from the  underlying event background in heavy-ion collisions. These techniques avoid, as far as possible, a reliance on theoretical assumptions as to the modification of the jet structure in Au-Au due to the interactions with the medium. First we needed to estimate the energy per unit area, $\rho$, of the background and its region-to-region fluctuations, $\sigma$, in a given event. Schematically, the reconstructed jet p$_{T}$ $\sim$ p$_{T}^{jet} + \rho A \pm \sigma \sqrt{A}$, where A is the jet cone area. In a central Au-Au event $\rho \sim$ 45 GeV for R=0.4 and $\sigma \sim$ 6-7 GeV when a p$_{T}$ cut of 0.2 GeV/c is applied to all  particles in the event. The scale of the fluctuations can  be significantly suppressed  by increasing the p$_{T}$ cut or decreasing the jet cone radius (resolution parameter), but only at the expense of the jet energy resolution and the loss of information about the softest part of the fragmentation~\cite{QMBruna}.

\begin{figure}[h] 
	\begin{center}
		\includegraphics[width=0.9\linewidth]{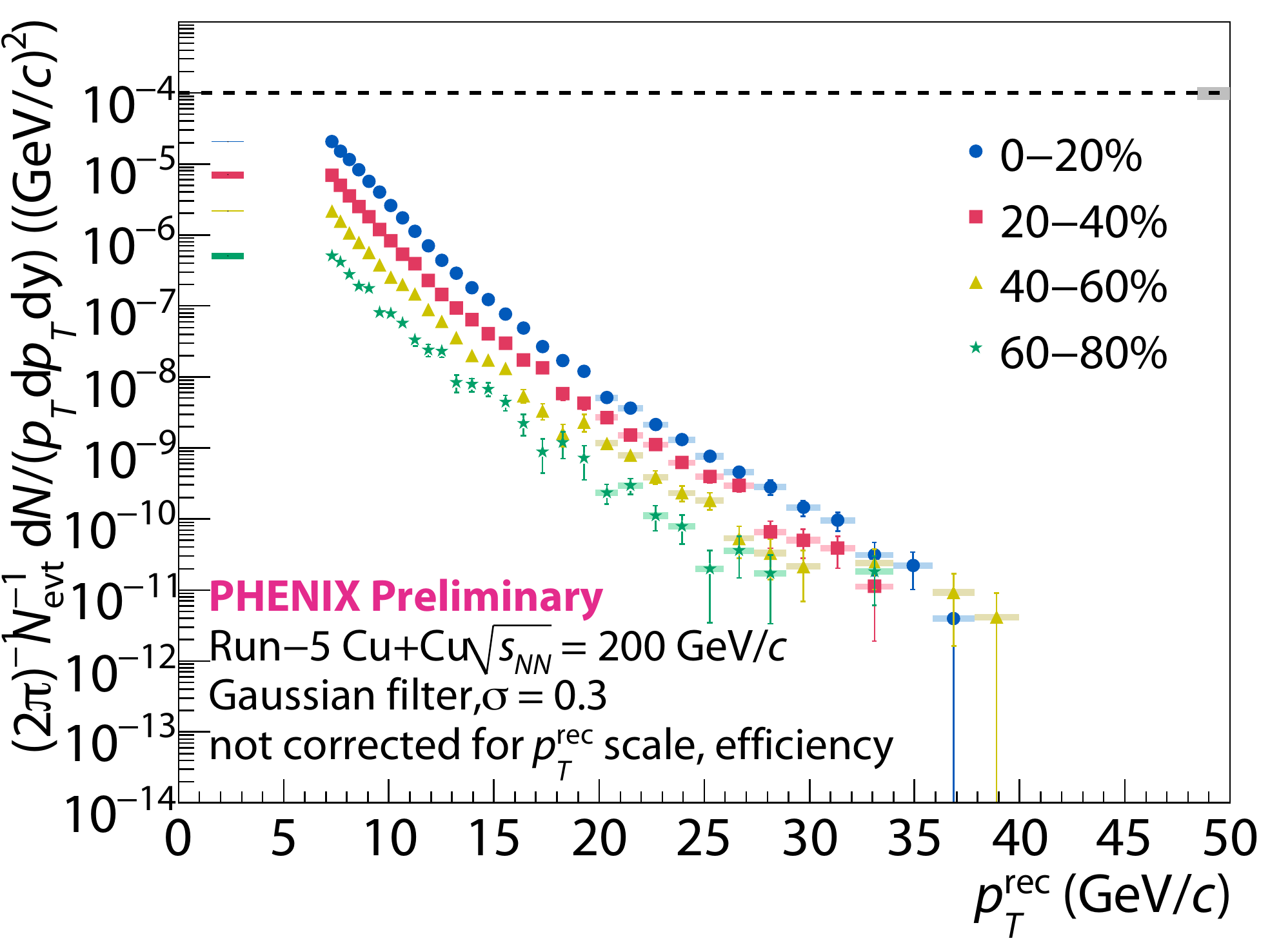}
		\caption{ Color online:   Fake rejected raw jet spectrum for Cu+Cu $\sqrt{s_{_{NN}}}$ = 200 GeV  collisions using
  $\sigma = 0.3$ Gaussian filter, $g_{\sigma_\mathrm{dis}} >
  17.8\,(\mathrm{GeV}/c)^2$ fake rejection. Figure from \cite{QMLai}.} \label{Fig:CuCuXSec}
	\end{center}
\end{figure} 

\begin{figure}[h] 
	\begin{center}
		\includegraphics[width=0.9\linewidth]{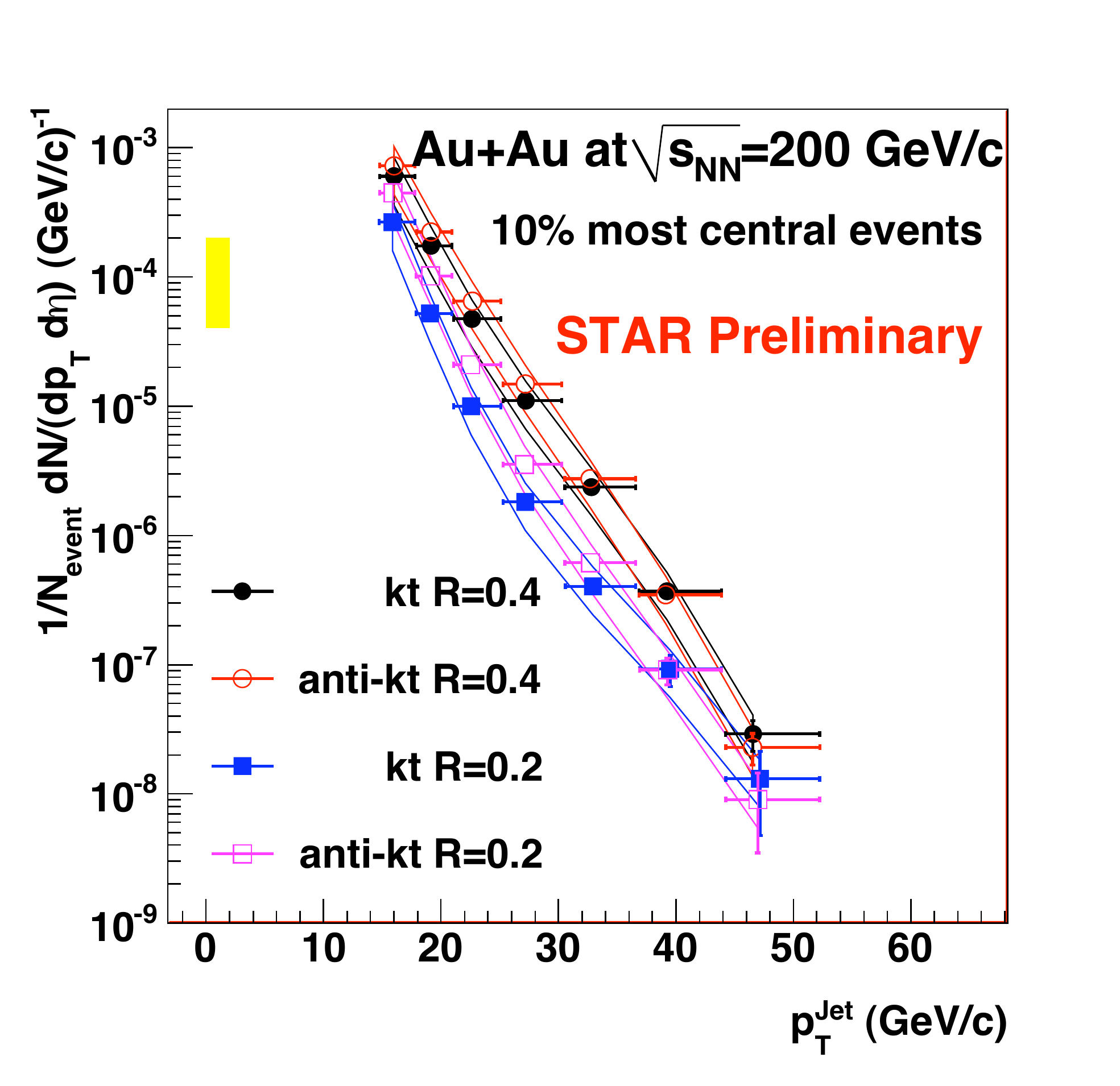}
		\caption{ Color online:  Cross sections for inclusive jet production in Au\,+\,Au collisions at $\sqrt{s_{NN}}=200~$GeV ($k_{T}$ and Anti-$k_{T}$, {\it R} = 0.2 and 0.4) $|\eta|<$1-R. The systematic uncertainty on the jet energy scale, shown as the vertical shaded band, is dominated by the
uncertainties on the electro-magnetic calorimeter calibration (5\%), unobserved jet energy (3\%), and charged
component momentum resolution (2\%). Solid curves represent the systematic
uncertainty on the jet yield in Au+Au due to background fluctuations. Figure from \cite{QMPloskon}.} \label{Fig:AuAuXSec}
	\end{center}
\end{figure}

Figures  \ref{Fig:CuCuXSec} and \ref{Fig:AuAuXSec}  show the inclusive jet cross-sections for various centralities in  Cu-Cu collisions \cite{QMLai}  and 0-10$\%$ central Au-Au collisions \cite{QMPloskon}  at $\sqrt{s_{_{NN}}}$ = 200 GeV respectively. These results, and those presented in~\cite{JoernHP}, represent the  first time that jets have been reconstructed in heavy-ion collisions and constitute a giant leap forward in RHIC physics. Figure\ref{Fig:CuCuXSec} is the raw reconstructed Cu+Cu jet spectrum for different centralities using a $\sigma$= 0.3 Gaussian filter. In this analysis fake jets are rejected by ensuring that   $g_{\sigma_\mathrm{dis}} (\eta, \phi) = \sum_{i \in \mathrm{fragment}}
  p_{T,i}^2 e^{-((\eta_i - \eta)^2 + (\phi_i -
    \phi)^2)/2\sigma_\mathrm{dis}}> 17.8\,(\mathrm{GeV}/c)^2$, $\sigma_{dis}$ = 0.1.  Fake jets are defined as the random clustering of energy that does not come from the fragmentation of a hard scattered parton  into a ``jet" candidate. A consistent power-law shape across all centralities is evident.

In Fig. ~\ref{Fig:AuAuXSec} the k$_{T}$ and Anti-k$_{T}$  jet finding algorithms are used. To obtain the  Au-Au inclusive jet yields the raw spectra are corrected for fake jets,  tracking inefficiencies, and the missed neutral energy, including K$_{L}^{0}$ and neutrons. The spectra are  also unfolded to account for jet energy resolution and background fluctuations.

\begin{figure}[h] 
	\begin{center}
		\includegraphics[width=\linewidth]{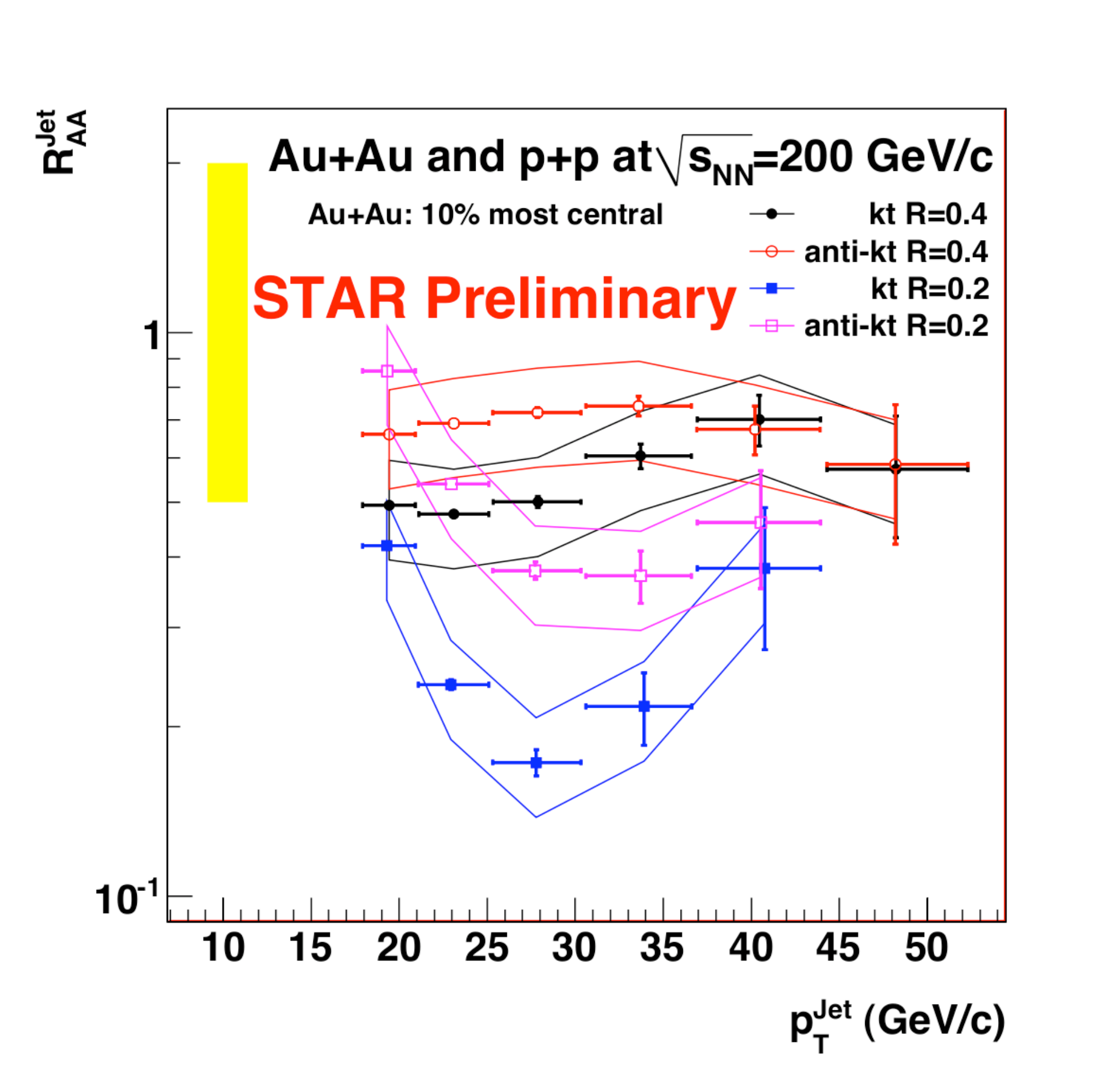}
		\caption{ Color online:  Jet R$_{AA}$ for  $\sqrt{s_{_{NN}}}$ = 200 GeV  0-10$\%$ Au-Au collisions for $k_{T}$ and
 		Anti-$k_{T}$, {\it R} = 0.2 and 0.4. Figure from \cite{QMPloskon}.} \label{Fig:JetRaa}
	\end{center}
\end{figure}

The  quenching effects of the medium discussed above will cause jet  fragmentation functions  in A-A collisions to be highly modified. The emitted gluon radiation results in numerous soft particles being produced enhancing the yield of low z particles. At the same time there will be a corresponding suppression of high z fragments. However, energy and momentum must be conserved even in the presence of jet quenching. Therefore if we can perform unbiased jet reconstruction the total initial partonic energy must be recovered, albeit with a re-distribution of the internal   jet structure. The nuclear modification factor is defined as:
\begin{eqnarray}
R_{AA} = \frac{Yield(A-A)}{Yield(p-p) \langle N_{coll} \rangle} 
\end{eqnarray}

where $ \langle N_{coll} \rangle$ is the average number of binary nucleon-nucleon collisions in the A-A collision, estimated via a Glauber calculation~\cite{Glauber}. If the jet reconstruction is unbiased then R$_{AA}$ = 1, if not all the energy is recovered then 
R$_{AA} <$1. Figure~\ref{Fig:JetRaa} shows the nuclear modification factor for jets in  0-10$\%$ Au-Au collisions for  R=0.4 and 0.2 using the $k_{T}$ and Anti-$k_{T}$ jet algorithms. It can be seen that the R$_{AA}$ for R=0.4 is  compatible with unity, within the large uncertainties. The R$_{AA}$   for R=0.2 for reconstructed p$_{T}^{jet} > $ 20 GeV/c on the other hand is markedly lower.  There are significant differences between the results for the  $k_{T}$ and Anti-$k_{T}$ algorithms. This is under investigation, but most likely arises from their different responses to the heavy-ion background~\cite{FastJet}.  

\begin{figure}[h] 
	\begin{center}
		\includegraphics[width=\linewidth]{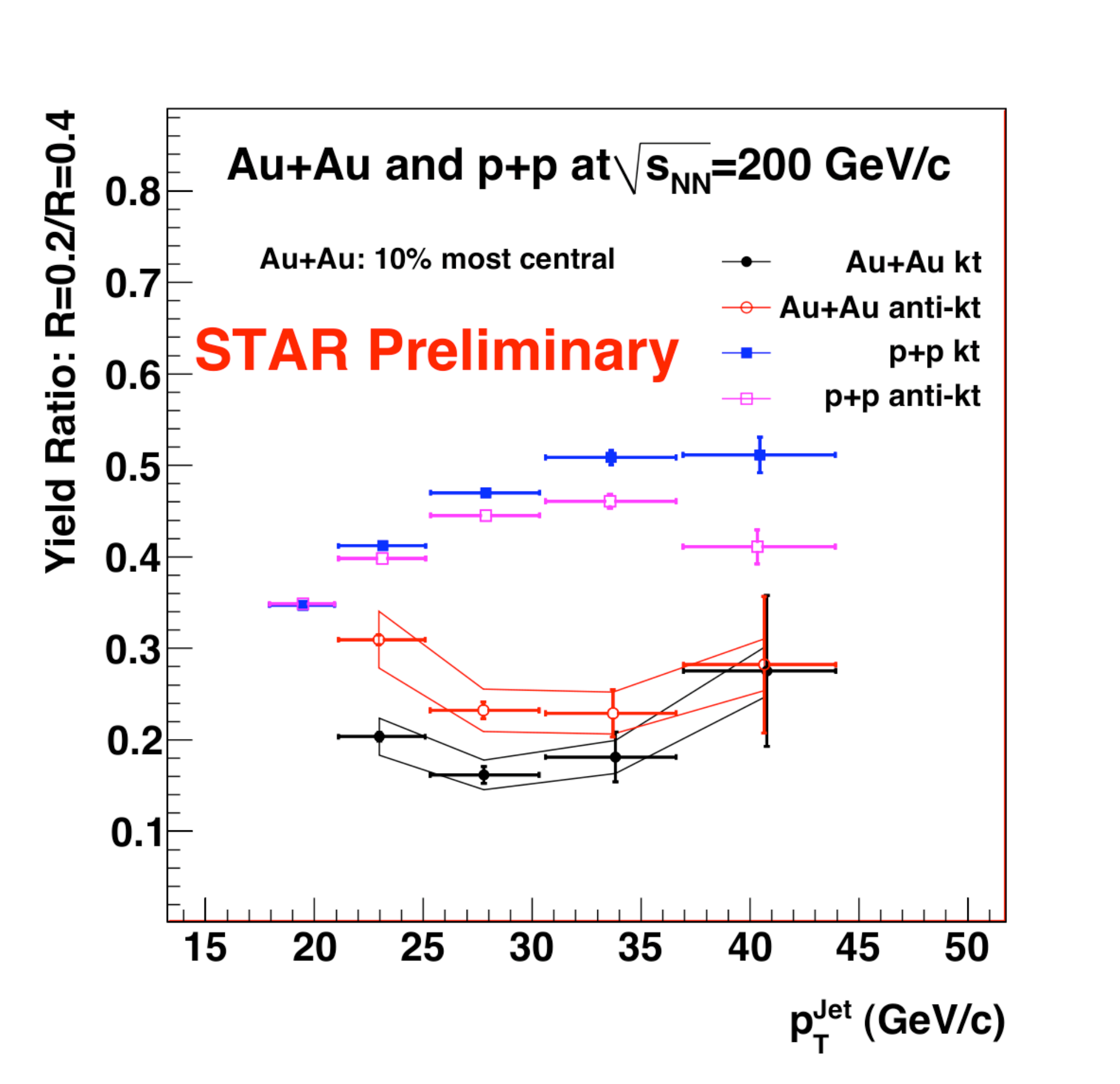}
		\caption{ Color online:  Ratios of  the R=0.2/R=0.4 inclusive jet
 		cross sections for $\sqrt{s_{_{NN}}}$ = 200 GeV  p-p and  0-10$\%$ Au-Au collisions  using $k_{T}$ and
 		Anti-$k_{T}$ algorithms. Figure from \cite{QMPloskon}.} \label{Fig:JetRadiiRatio}
	\end{center}
\end{figure} 

Figure~\ref{Fig:JetRadiiRatio} shows the ratio of inclusive jet cross-section for R=0.2/R=0.4 jet resolution parameters for p-p and Au-Au data. This ratio increases with reconstructed jet energy for p-p collisions, as expected, due to the focussing of the jet fragmentation with increasing partonic energy.  However, the Au-Au data shows a different trend, the ratio is approximately flat.  In this representation, there is no difference in the results from the k$_{T}$ and Anti-k$_{T}$ jet algorithms, supporting the notion that the differing results observed in  Fig.~\ref{Fig:JetRaa} are indeed due to their  behaviors in the presence of the large  background in Au-Au events.

\subsection{Jet-hadron Correlations}
\begin{figure}[h] 
	\begin{center}
		\includegraphics[width=\linewidth]{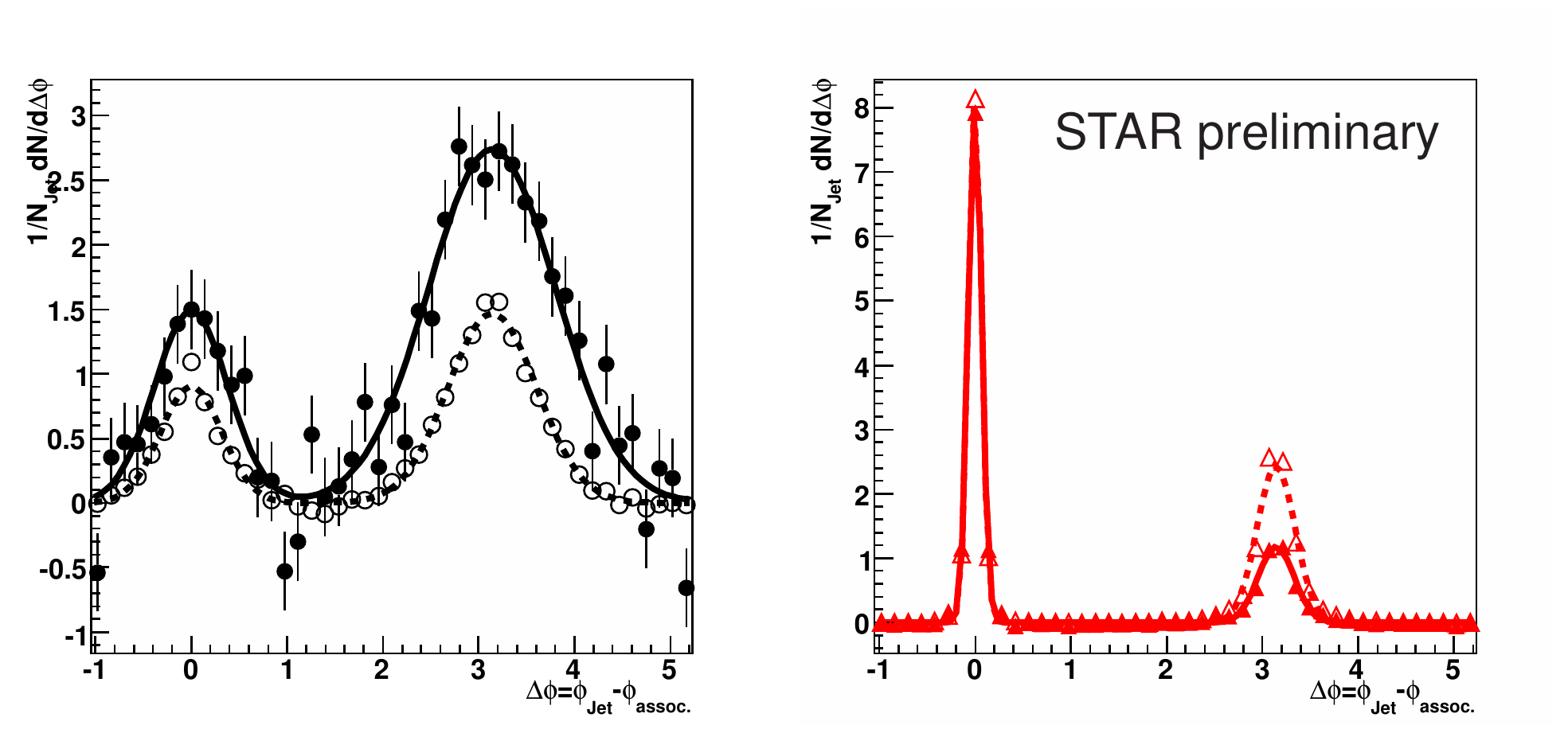}
		\caption{ Color online: Jet-hadron $\Delta\phi$ correlations.  Reconstructed jet p$_{T} >$  20 GeV/c, using Anti-k$_{T}$ algorithm, R=0.4 and p$_{T} >$ 2 GeV/c for all charged particles and towers. Left: Associated particles 0.2 $<$ p$_{T}$ $<$ 1.0 GeV/c and Right:   Associated particles  p$_{T} >$ 2.5 GeV/c. The open symbols are for p-p collisions and the closed symbols for 0-10$\%$ Au-Au $\sqrt{s_{_{NN}}}$ = 200 GeV  collisions. Figure from \cite{AGSPutschke}.} \label{Fig:JetHadron}
	\end{center}
\end{figure}

Figure~\ref{Fig:JetRaa} and  ~\ref{Fig:JetRadiiRatio}  indicate that the presence of the medium results in a sizeable de-focussing of the jet resulting in significant  amounts of the initial partonic energy being lost. To study this further we use jet-hadron correlations. First the Anti-k$_{T}$ algorithm with a resolution parameter of R=0.4 is run on  events where only charged tracks and electromagnetic calorimeter towers with p$_{T}  >$ 2 GeV/c are considered. If a jet is reconstructed with  p$_{T} >$  20 GeV/c the  $\Delta\phi$  of  charged particles is calculated  with respect to the jet axis.  This jet-hadron correlation allows the study of the away-side jet  even in the case that this di-jet partner would not be reconstructed via  standard jet algorithms. The resulting correlations are shown in Fig.~\ref{Fig:JetHadron} for associated  hadrons with 0.2 $<$ p$_{T}$   $<$ 1.0 GeV/c (left panel) and  p$_{T}$   $>$ 2.5 GeV/c (right panel). It can be seen (right panel) that for high p$_{T}$ associated particles the away side jet is dramatically suppressed in Au-Au compared to p-p collisions. However, for the low p$_{T}$ associated particles the Au-Au data is not only strongly enhanced but also significantly broader, with significant enhancement beyond a radius of 0.4. Using this type of analysis will allow us to establish the scale of the ``out-of-cone"  radiation emitted by the parton, which is not possible with conventional jet reconstruction. This result represents the first direct measurement of the ``modified fragmentation" of jets  in the presence of the sQGP. 

\subsection{Summary of Jet Studies}

Jet studies are well under way at RHIC. The p-p reference measurements are well described by theoretical calculations, indicating that we have a calibrated probe to turn on the heavy-ion medium. The underlying event in p-p collisions is small and the measurements indicate that there is little to no large angle initial and final state radiation. In addition the cold nuclear matter effects, as measured via d-Au collisions, are minimal. 

The  background due to the hadronization of the sQGP created at RHIC, while significant,  can be statistically subtracted from the reconstructed jets. There is strong evidence that the  partons resulting from  initial high Q$^{2}$ scatterings lose significant energy as they traverse the colored medium. Their jet profile is strikingly broadened and the hadrons produced have much softer p$_{T}$ on average than jets fragmenting in vacuum, as determined via p-p collisions at the same beam energy.

\section{ Local Strong Parity Violation}

Although parity violation in weak interactions was first  observed in 1975~\cite{PVWeak}, until recently it was thought to be conserved in strong interactions. Modern QCD calculations now allow for such parity violation but to date it has not been observed. It has  been suggested that in relativistic heavy-ion collisions parity-odd domains may be created which correspond to  non-trivial topological solutions of the  QCD vacuum thus  violating parity and time-reversal symmetries locally~\cite{LPV}.  In non-central collisions,  significant orbital angular momentum and an extremely  large magnetic field, albeit short lived, are created when the two nuclei collide. The non-zero topological charge domains produce different numbers of  positive and  negative quarks in each domain. The interaction of these particles with the strong B-field results in a  charge separation of quarks along the systems orbital momentum axis ~\cite{PV1, PV2, PV3}.  A clear division of charge along the angular momentum vector would therefore be evidence of local strong parity violation. 

In any given event this charge separation along the  angular momentum vector may be described by
\begin{eqnarray*}
\frac{dN_{\pm}}{d\phi}  \approx ( 1 + 2v_{1}cos(\Delta\phi) + 2v_{2}cos(2\Delta\phi)+...+\\
 2a_{\pm}sin(\Delta\phi)+...),
\end{eqnarray*}

where $\Delta\phi$ is the azimuthal angular difference  of each particle with respect to the reaction plane. The reaction plane is defined by the beam axis and the line joining the centers of the two colliding nuclei. $v_{1}$ and $v_{2}$ etc are the coefficients that account for the so-called directed and elliptic flow~\cite{Flow}. The ``$a$" parameters, where $a_{+}$ = - $a_{-}$, measure the parity violating effect. However, since this is a spontaneous effect  the  sign of $a_{+}$ (and hence $a_{-}$)   is random in each event, when averaged over many events this parity odd observable therefore yields, $\langle a_{+} \rangle $ = $\langle a_{-} \rangle $=  0.  

It was therefore suggested~\cite{PVExp} to instead  detect  
\begin{eqnarray*}
\langle a_{\alpha} a_{\beta} \rangle  \approx \langle cos(\phi_\alpha + \phi_{\beta} - 2\Psi_{RP} )\rangle\end{eqnarray*}
where $\alpha$ and $\beta$ are the sign of the electric charge of the particles sampled and $\Psi_{RP} $ is the reaction plane angle. Experimentally this can be  measured via a three particle correlation since

\begin{eqnarray*}
 \langle cos(\phi_\alpha + \phi_{\beta} - 2\Psi_{RP} )\rangle
 = \langle cos(\phi_\alpha + \phi_\beta - 2\phi_c) \rangle/v_{2,c}
\end{eqnarray*}
 
 where  the three particles are labelled $\alpha, \beta$, and c. It is assumed that the only common correlation between particles $\alpha, \beta$ and c is via the reaction plane, and particle c is used only to determine the angle of the event plane. 
 
 Assuming  this  correlation to be only sensitive to the parity violation effect, and ignoring any final state interactions with the medium, then  $\langle a_{+} a_{+} \rangle$ = $\langle a_{-} a_{-} \rangle$ = -$\langle a_{-} a_{+} \rangle$ $>$ 0.  This correlator should also scale inversely with the number of particles produced, assuming the domain sizes stay constant.  However, it  seems extremely likely that there are interactions with the  medium the predicted  effect of these interactions is a suppression of the back-to-back 3 particle correlation (one of the charged pairs  has to make its way through the medium) resulting in  $\langle a_{+} a_{+} \rangle$ = $\langle a_{-} a_{-} \rangle$  $>>$ $\langle a_{-} a_{+} \rangle$. 
 
\begin{figure}[h]
  \begin{center}
    \includegraphics[width=\linewidth]{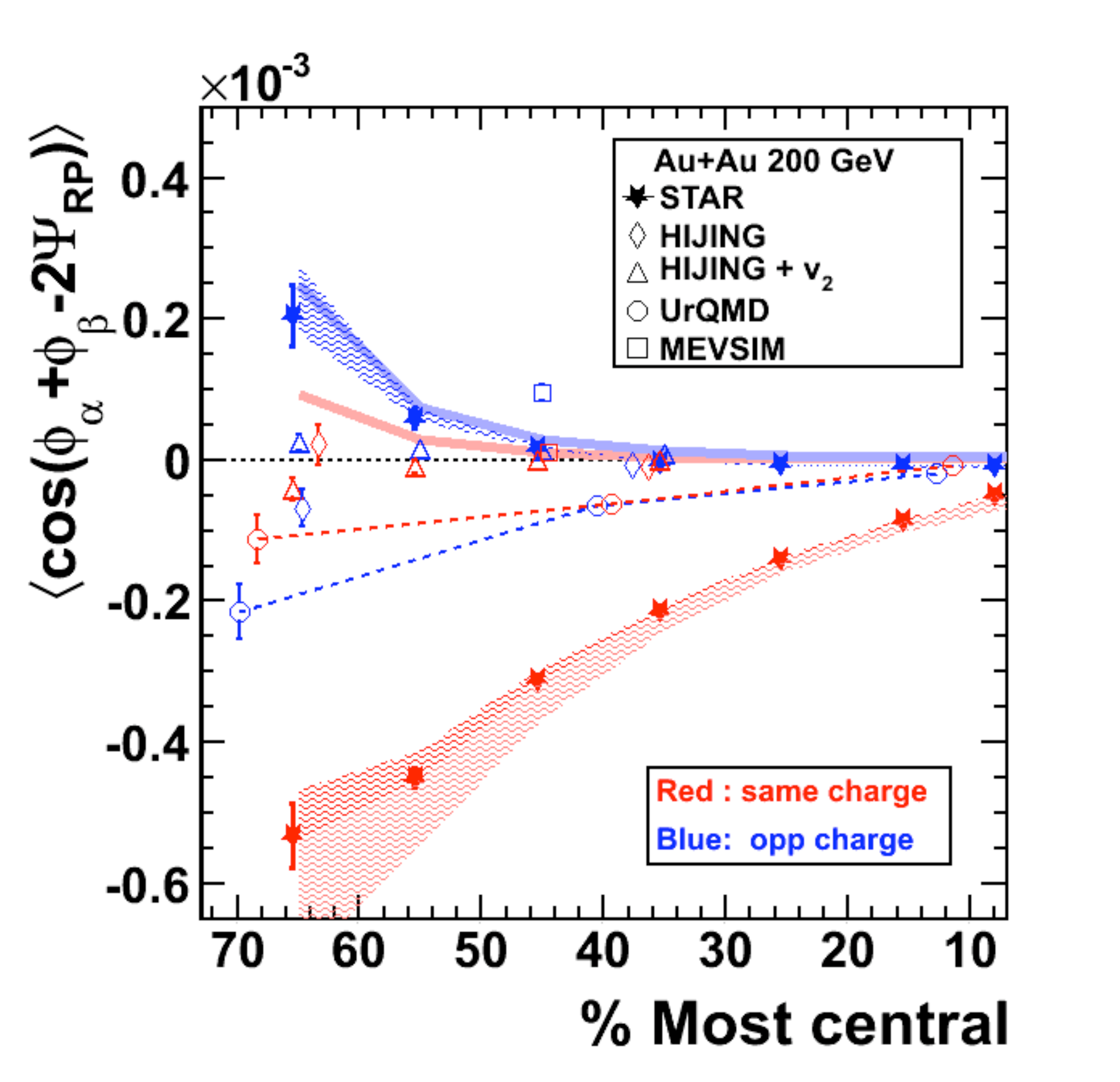}
  \end{center}
  \vspace{-1pc}
  \caption{Color Online: $\langle cos(\phi_{\alpha} + \phi_{\beta} - 2\Psi_{RP}) \rangle$ calculated from data from Au+Au collisions at  $\sqrt{_{s_{NN}}}$= 200 GeV collisions by the  STAR experiment. Also shown are the same calculations from the HIJING (with and without elliptic flow ``after burners"), UrQMD, and MeVSIM event generators. From~\cite{STARPV}.}  \label{Fig:PV}
\end{figure}

 Figure ~\ref{Fig:PV} shows the correlations for Au-Au collisions at $\sqrt{s_{_{NN}}}$ = 200 GeV as a function of centrality for same-sign and opposite-sign charge values of $\alpha$ and $\beta$. As expected for the parity violation signal, the  absolute magnitude of the opposite-charge correlations are smaller than the same-charge correlations, and the sign of the correlations are opposite. The predicted inverse dependence on centrality is  observed. 
 
 However, since this measure is now parity-even it is possible for other effects to produce a non-zero signal. Examples of such physics backgrounds considered are coulomb effects, the decay of resonances, jet fragmentation and collective motion of the bulk.  One technique used to investigate such possibilities was the calculation of the three particle correlation described above for the HIJING~\cite{Hijing}, UrQMD~\cite{UrQMD} and MeVSim~\cite{MeVSim} heavy-ion event generators. The results are also shown in Fig.~\ref{Fig:PV} MeVSim only includes  correlations due to resonance decays and overall elliptic flow, while HIJING and UrQMD are real physics models of the collisions and therefore include correlations
due to many physical processes.  It can be seen that none of these generators can reproduce the data. So far no non-parity violating signals have been found, via our systematic checks, that can explain the magnitude, sign and centrality dependence of both the like- and opposite-sign correlations. For further details of these studies see~\cite{PVChecks}.

\section{The Next Frontiers}
Current relativistic heavy-ion experiments  are poised to explore two exciting new frontiers: 1) the new energy regime made accessible  with the turn-on of the Large Hadron Collider (LHC), scheduled for November 2009 and 2) a Beam Energy Scan at RHIC set to start in 2010.

\subsection{Beam Energy Scan}

\begin{figure}[h]
  \begin{center}
    \includegraphics[width=0.9\linewidth]{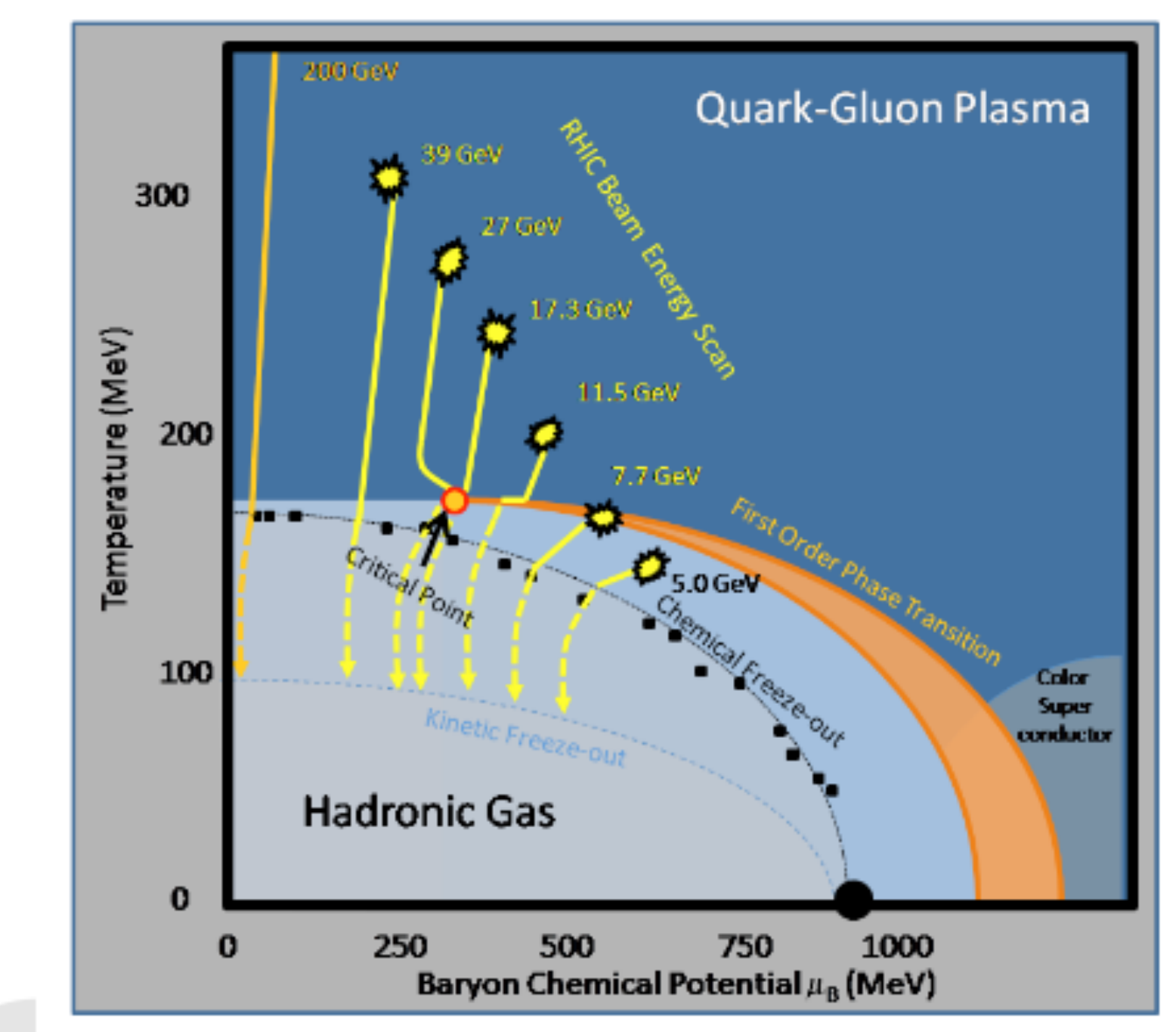}
  \end{center}
  \vspace{-1pc}
  \caption{Color Online: A sketch of the  phase diagram of nuclear matter. The location of the critical point is  suggestively placed  within the RHIC BES range, not estimated from any theoretical calculation. The black closed circles are current heavy-ion experimental calculations of the chemical freeze-out temperature, T$_{{\rm ch}}$, and $\mu_{B}$ based on statistical model fits to the measured particle ratios. The yellow curves show the estimated trajectories of the  collision energies proposed for the RHIC BES. }  \label{Fig:PhaseDiagram}
\end{figure}

Theory calculations can be used to predict  how transitions to a sQGP depend on the baryon chemical potential, $\mu_{B}$, and temperature, T. At low $\mu_{B}$ and high T a cross-over transition is predicted, while at high $\mu_{B}$  and low T the transition is of  first order. Hence, at intermediate values, a critical point should occur (see for example \cite{BES}). Experimentally we can vary these initial conditions by altering the beam energy. At RHIC we can explore the range from $\sqrt{s_{_{NN}}}$ = 5-200 GeV covering the $\mu_{B}$ range of  500-20 MeV. The phase diagram of nuclear matter is depicted in Fig.~\ref{Fig:PhaseDiagram}   The yellow curves in the figure are the trajectories expected for the various collision energies selected for the beam energy scan (BES).  After thermalization, the system rapidly cools and expands before passing through the phase transition from the sQGP to a hadronic gas. Inelastic  hadronic re-scatterings  continue until chemical freeze-out is reached, at which point the ratio of particle species in locked in. Elastic interactions continue until the system passes through kinetic freeze-out, at which point the hadrons free stream to the detectors. Theoretical calculations are not currently able to determine the exact location of the critical point. It has  therefore been suggestively placed in the centre of the range  of the RHIC BES. A RHIC beam energy scan will allow us to  explore the  QCD phase diagram close to the QGP-hadron gas boundary  and locate such  key ``landmarks"  as the  critical point. Establishing the existence of  this critical point would be a seminal step forwards for QCD physics. The BES would also allow us to determine the collision energy at which a sQGP is no longer made, by identifying the $\sqrt{s_{_{NN}}}$ at which the signals associated with the plasma are no longer observed. Finally the BES could be used to further study the potential local parity violation signal discussed above. If it is a parity violation signal the effect should be stronger  at lower  collision energies, as the time integral of  the magnetic field is larger. The signal should however not persist at collision energies where no sQGP signals are formed since the parity violating effect is dependent on  de-confinement. 

\subsection{The Large Hadron Collider}
A new energy window  for particle physics is set to open at the end of 2009 when the LHC commences operation. First Pb-Pb collisions at $\sqrt{_{s_{NN}}}$= 5.5 TeV are currently scheduled for the end of 2010. This order of magnitude increase in collision energy  will allow the realm of jet and heavy flavor physics, just starting to be exploited at RHIC, to be fully explored with high statistics. The initial conditions will be  hotter and denser, and the medium produced is predicted to be larger and longer lived than that currently being studied by STAR and PHENIX~\cite{LHCPredictions}, see table~\ref{Table:LHCRHIC} It will be very interesting to study if the sQGP created  is the same or if other,  unexpected properties, emerge. Currently there is only one dedicated heavy-ion experiment, ALICE~\cite{Alice}, but both CMS~\cite{CMS} and ATLAS~\cite{Atlas} have active heavy-ion groups. 

In addition, the RHIC accelerator is currently being upgraded (RHIC-II) for higher luminosity running via  stochastic cooling of the beams. While the hard probes studies will certainly be dominated by the LHC results, RHIC will stay competitive in the low to intermediate p$_{T}$ ranges due to its longer running time per year and higher luminosity after the RHIC-II upgrade. By combining the RHIC and LHC data we will therefore be able to make studies of the physics of heavy-ion phenomena in unprecedented detail from 0.005-5.5 TeV. 

\begin{table}[h]
\small
    \caption{Comparison of  measured RHIC quantities to those expected at the LHC. The physical quantities of the medium are for central Au-Au (RHIC) or Pb-Pb (LHC) collisions at maximum beam energy.  The beam measures are for nominal annual heavy-ion running.} 
\begin{center}
\begin{tabular}{|clc|c|} \hline
 & RHIC-II &       LHC  \\   \hline
Collision energy in A-A         & 5-200 GeV & 5.5 TeV  \\
Initial temperature    &   $\sim$2T$_{c}$ & $\sim$4T$_{c}$  \\
Initial Energy density     & 5 GeV/fm$^{3}$ &  15-60 GeV/fm$^{3}$  \\   
Lifetime &  2-4 fm/c  & $>$ 10 fm/c \\
Data taking period & 12 weeks & 4 weeks \\
Ave. A+A L (cm$^{-1}$s$^{-1}$) & 5x10$^{27}$ & 5x10$^{26}$ \\
Integrated L (50$\%$ uptime) & 20nb$^{-1}$ & 500$\mu$b$^{-1}$ \\

\hline
\end{tabular}
\end{center}
\end{table}\label{Table:LHCRHIC}

\section{Summary}

There are many new and exciting results being produced at RHIC. 

While a complete theoretical understanding of these data is still not established, there are many predictions left for heavy-ion experiments to test. Key questions that remain not fully answered are 
\begin{itemize}
\item How do partons interact with the sQGP and what are the exact mechanisms of their energy loss?
\item What are the properties of the medium produced and how they change with differing initial conditions?
\item What and where are the key landmarks on the QCD phase diagram of nuclear matter?
\item Can the possible signal of  local strong parity violation, seen in RHIC collisions and a fundamental prediction of QCD, be explained by any other more mundane signals.
\end{itemize}

RHIC and the LHC have exciting new programs starting in 2010 which will attempt to answer these, and other questions. The results from the experiments based at these colliders will provide new and complimentary data with which to explore QCD under extreme conditions.

     \vspace{5\baselineskip}

%%%%%%%%%%%%%%%%%%%%%%%%%%%%%%%%%%
%% thebibliography environment %%
%%%%%%%%%%%%%%%%%%%%%%%%%%%%%%%%%

%%%%%%%%%%

\end{document}